\documentclass[times,twocolumn,final]{elsarticle}

\usepackage{cag}
\usepackage{framed,multirow}

\usepackage{amssymb}
\usepackage{latexsym}

\usepackage{url}
\usepackage[table]{xcolor}
\definecolor{newcolor}{rgb}{.8,.349,.1}
\definecolor{author}{RGB}{0,128,128}
\usepackage{enumitem}

\usepackage{hyperref}
\usepackage[switch,pagewise]{lineno} 

\usepackage{booktabs}
\usepackage{amsmath}
\usepackage{pifont}
\usepackage{xspace}
\newcommand{\ie}{\emph{i.e.}\xspace}
\newcommand{\eg}{\emph{e.g.}\xspace}

\newcommand{\myparagraph}[1]{\smallskip\noindent\textbf{#1}\xspace}

\newcommand{\systemname}{\emph{STRive}\xspace}

\usepackage{tcolorbox}
\tcbset{colback=gray!5!white, colframe=gray!80!black, fonttitle=\bfseries}

\newtcolorbox[auto counter, number within=section]{chatbox}[1][]{%
  colback=gray!5!white, colframe=gray!80!black, float=t!,
  fonttitle=\bfseries, title=Conversation~\thetcbcounter: #1, #1,
  label={chat:#1}
}

\tcbset{colback=gray!5!white, colframe=gray!80!black, fonttitle=\bfseries}

\newtcolorbox[auto counter, number within=section]{chatboxx}[1][]{%
  colback=gray!5!white, colframe=gray!80!black, float=h!,
  fonttitle=\bfseries, title=Prompt~\thetcbcounter #1, #1,
  label={chatx:#1}
}

\journal{Computers \& Graphics}

\begin{document}

\verso{Preprint Submitted for review}

\begin{frontmatter}

\title{STRive: An association rule-based system for the exploration of spatiotemporal categorical data}%

\author[1]{Mauro \snm{Diaz}\corref{cor1}}
\cortext[cor1]{Corresponding author:}
\emailauthor{dany.espino@fgv.br}{Mauro Diaz}
    
\author[1]{Luis \snm{Sante}}

\author[1]{Joel \snm{Perca}}

\author[2]{João \snm{Victor da Silva}}

\author[2]{Nivan \snm{Ferreira}}
\fntext[fn1]{Footnote 1.}  

\author[1]{Jorge \snm{Poco}}
\address[1]{Fundação Getulio Vargas. Praia de Botafogo, 190 - Botafogo, Rio de Janeiro - RJ, 22250-145, Brazil}
\address[2]{Universidade Federal de Pernambuco. Av. Jornalista Aníbal Fernandes, s/n – Cidade Universitária, Recife, Pernambuco, Brazil}

\received{\today}

\begin{abstract}
Effectively analyzing spatiotemporal data plays a central role in understanding real-world phenomena and informing decision-making. Capturing the interaction between spatial and temporal dimensions also helps explain the underlying structure of the data. However, most datasets do not reveal attribute relationships, requiring additional algorithms to extract meaningful patterns. Existing visualization tools often focus either on attribute relationships or spatiotemporal analysis, but rarely support both simultaneously.
In this paper, we present \systemname{} (SpatioTemporal Rule Interactive Visual Explorer), a visual analytics system that enables users to uncover and explore spatial and temporal patterns in data. At the core of \systemname{} lies Association Rule Mining (ARM), which we apply to spatiotemporal datasets to generate interpretable and actionable insights. We combine ARM with multiple interactive mechanisms to analyze the extracted relationships. Association rules serve as interpretable guidance mechanisms for visual analytics by highlighting the meaningful aspects of the data that users should investigate.
Our methodology includes three key steps: rule generation, rule clustering, and interactive visualization. \systemname{} offers two modes of analysis. The first operates at the rule cluster level and includes four coordinated views, each showing a different facet of a cluster, including its temporal and spatial behavior. The second mode mirrors the first but focuses on individual rules within a selected cluster.
We evaluate the effectiveness of \systemname{} through two case studies involving real-world datasets\,---\,fatal vehicle accidents and urban crime. Results demonstrate the system's ability to support the discovery and analysis of interpretable patterns in complex spatiotemporal contexts.
\end{abstract}

\begin{keyword}
\KWD Spatiotemporal visualization, Association rule mining, Pattern extraction, Visual Analytics system.
\end{keyword}

\end{frontmatter}


\section{Introduction}
\label{sec:introduction}

Spatiotemporal data mining continues to pose significant challenges in real-world scenarios because relationships between attributes often lack evident structure~\cite{hamdi2022spatiotemporal,10.1145/3696661,zhang2024survey}.
Due to their inherent complexity, visualization methods aimed at exploring spatiotemporal phenomena increasingly benefit from user-guided mechanisms, which have gained prominence in visual analytics environments~\cite{perezmessina2024enhancing}.
A notable guidance mechanism is \textit{Association Rule Mining (ARM)}, which is widely employed to reveal meaningful relationships within datasets~\cite{agrawal1993mining}. 
ARM systematically extracts associations, clearly highlighting co-occurrences and conditional relationships among variables.

Association rules provide effective suggestions or initial exploration points within visual analytics workflows, enabling users to focus analyses on specific data aspects. 
Each rule explicitly defines relationships through \textit{``if-then''} statements, clarifying the interactions between variables.
Instead of initiating analyses from scratch or employing manual queries, analysts leverage these rules as preliminary hypotheses, accelerating their discovery processes.
Furthermore, association rules offer greater interpretability than opaque models, such as neural networks or ensemble methods, allowing analysts to comprehend underlying patterns without requiring intricate explanations.

Although ARM demonstrates extensive applicability across various domains, its integration within spatiotemporal visual analytics faces considerable challenges. 
Generating large-scale association rules introduces computational complexities, and effectively visualizing these rules constitutes a significant hurdle in spatiotemporal contexts.

In this work, we present a novel methodology to address these challenges and uncover interpretable patterns in spatiotemporal datasets.
The approach applies association-rule mining to automatically detect significant patterns, which are then systematically organized into meaningful clusters. 
We implemented the methodology in a visual analytics system, \systemname, which provides multiple coordinated views that facilitate the interactive exploration of these patterns using well-established visualizations.
%
%
The system enables a structured drill-down analysis, beginning with general trends and progressing toward detailed insights. 
\systemname effectively directs analysts' attention to critical areas of interest, enabling more focused and efficient exploration. 
\systemname's
interactive visualizations enable users to thoroughly investigate rule compositions, temporal dynamics, pattern comparisons, and spatiotemporal occurrences. 
By combining powerful computational methods with intuitive visual analytics, \systemname significantly enhances analysts' ability to interpret and gain actionable insights from spatiotemporal data. 
%
%
%
Our contributions are as follows:

\begin{itemize}[noitemsep]
    \item A methodology using established algorithms for identifying and clustering temporal patterns. 
    \item \systemname, an interactive visual analytics tool that significantly improves exploration and interpretation of spatiotemporal datasets. 
    \item Two comprehensive case studies using real-world datasets that demonstrate significant findings, accompanied by clear and insightful explanations.
\end{itemize}

\section{Related Works}
\label{sec:related}

Analyzing spatiotemporal data patterns is necessary for public safety, transportation, epidemiology, and environmental monitoring~\cite{hamdi2022spatiotemporal}. 
Visual analytics systems supporting this analysis typically use coordinated interactive views. For comprehensive reviews of these visualization techniques, see Liu~et~al.~\cite{liu2016visualizing} and Mota~et~al.~\cite{mota2022comparison}. 
Unlike purely user-driven exploration, our method automatically identifies interpretable patterns first and then uses interactive visualizations for detailed analysis. 
This section reviews guidance in visual analytics, association rule visualization, and hypergraph representations.

\subsection{Guidance in Visual Analytics}


Guidance in visual analytics is ``a computer-assisted process that actively resolves a knowledge gap encountered by users during an interactive visual analytics session''~\cite{ceneda2016characterizing}. 
According to Collins~et~al.~\cite{collins2018guidance}, guidance supports analysts in identifying and managing data patterns. 
They propose approaches for implementing guidance, such as facilitating pattern extraction\,---\,visually through effective representations or computationally via suitable algorithms. 
They also emphasize guidance in pattern evaluation, considering properties like frequency, intensity, and rate of change. Several studies have introduced strategies for selecting~\cite{perezmessina2024enhancing, sperrle2023lotse, wenkai2023providing} and evaluating~\cite{collins2018guidance} guidance techniques, underscoring their practical importance.

Numerous visual analytics systems have incorporated guidance mechanisms. 
Doraiswamy~et~al.~\cite{doraiswamy_using_2014} propose a method that uses topological features to identify events in scalar time-varying data automatically. 
Their visual interface guides users toward noteworthy events and supports similarity searches. Valdivia~et~al.~\cite{valdivia_wavelet-based_2015} represent data as graphs and apply graph wavelet transforms to detect regions with significant variation, identifying abrupt changes as important events. 
Liu~et~al.~\cite{liu_tpflow_2019} developed a workflow for pattern discovery and comparison, utilizing tensor decomposition to recommend optimal data slices and highlight patterns automatically.

Although these approaches successfully utilize algorithms to guide exploration, the complexity of the underlying methods may require users to exert additional effort to interpret the extracted patterns.

\subsection{Association Rules in Visual Analytics}

Association rules summarize interpretable relationships within data and serve as effective guidance mechanisms for decision-making. 
Varu~et~al.~\cite{varu2022armatrix} present a framework using an item-to-rule matrix visualization. Items form rows, rules form columns, and glyphs indicate antecedent-consequent relationships. 
Metrics are visually represented through histograms aligned with rows and columns. 
Menin~et~al.~\cite{menin2021arviz} utilize linked visualizations, including scatterplots for rule metrics, chord diagrams for item relationships, and association graphs for exploratory item analysis. 
Wang~et~al.~\cite{wang2018association} employ association rules to analyze spatiotemporal climate data, discretizing continuous attributes for effective analysis. 
Their approach integrates spatial context, enriching the analysis beyond attribute relationships alone.

Despite advances, existing visualizations for association rules often lack sufficient contextual visualization, leading to information loss. 
Fister~et~al.~\cite{fister2023comprehensive} provide a detailed critique of these limitations. 
In contrast, our system STRive addresses this limitation by embedding association rules within a rich spatiotemporal framework. 
It enables users to analyze the relationships between attributes and \textit{when} and \textit{where} these patterns emerge.

\subsection{Hypergraph Representations}

Association rules can be modeled using hypergraphs. In this representation, items are represented as nodes, and rules are defined as hyperedges that connect multiple nodes. 
Visualizing such structures is challenging. 
Valdivia~et~al.~\cite{valdivia2021analyzing} propose a matrix-like visualization to display dynamic hypergraphs, with nodes represented as rows and hyperedges as vertical lines connecting relevant nodes across time. 
Oliver~et~al.~\cite{oliver2024scalable} develop polygon-based hypergraph visualizations using iterative simplification to optimize layouts. 
Additional node-link, timeline-based, and matrix-based visualization approaches are reviewed in Fischer~et~al.~\cite{fischer2021towards}.

Our work draws particularly from Valdivia~et~al.~\cite{valdivia2021analyzing}, adopting their approach of representing ordered sets of hyperedges (rules) over time. 
We extend this method by encoding additional contextual data to support comprehensive analysis.
\section{Background}
\label{sec:background}

We now briefly present key concepts in Association Rule Mining (ARM). For a more detailed discussion on this topic, see Zhang~and~Zhang~\cite{zhang2002association} and Nath~et~al.~\cite{nath2013incremental}.
Agrawal et al.~\cite{agrawal1993mining} first introduced ARM as a data mining method to uncover frequent associations between variables in a dataset.
Let $\mathcal{D}$ denote the dataset, and capital letters (\eg, $X, Y$) represent subsets of $\mathcal{D}$.
Following Kisilevich et al.~\cite{Kisilevich2010}, we focus on spatiotemporal (ST) event datasets. Each entry in the dataset is a point in space-time, defined as $p=(location_p, t_p, attrib_p)$, where $attrib_p$ is a vector of additional attributes.
To apply ARM, we assume all components, including location and time, are categorical. 
When necessary, we discretize geographic coordinates into administrative regions (\eg, countries or states) and time into intervals such as months or years.

ARM expresses relationships as association rules in the form $r:= X \Rightarrow Y$, where $X$ is the antecedent ($Ante(r)$) and $Y$ is the consequent ($Cons(r)$).
A rule states that if $X$ appears, then $Y$ is likely to follow.
For instance, in a sales dataset, the rule $\{\text{Bread}, \text{Coffee}\} \Rightarrow \{\text{Milk}\}$ suggests that customers who buy bread and coffee often also purchase milk.
Although originally applied to market basket analysis~\cite{agrawal1993mining}, researchers now use ARM in other domains such as traffic accident analysis~\cite{MOMENIKHO2021101278}, electronic health records~\cite{LAKSHMI2017290}, and microblogging text analysis~\cite{9078743}. 

The ARM process begins by identifying \emph{frequent itemsets}, \ie, values that commonly appear together. 
Algorithms like Apriori~\cite{agrawal1994fast} and FP-Growth~\cite{han2000mining} perform this task. 
From these \textit{itemsets}, the algorithm generates association rules by splitting the itemsets into antecedent and consequent parts.
To evaluate the quality of the generated rules, ARM uses several metrics:
For example:

\begin{itemize}[nosep,leftmargin=*]
    \item \textbf{Support}~\cite{agrawal1993mining} measures how often an itemset appears in the dataset: $sup(X) = |X| / |\mathcal{D}|$. For a rule, $sup(X \Rightarrow Y) = sup(X \cap Y)$.
    \item \textbf{Confidence}~\cite{agrawal1993mining} estimates the probability of $Y$ appearing when $X$ is present: $conf(X \Rightarrow Y) = |X \cap Y| / |X|$.
    \item \textbf{Lift}~\cite{10.1145/253260.253325} measures the strength of the association by comparing the joint probability of $X$ and $Y$ to their expected probability if independent: $lift(X \Rightarrow Y) = sup(X \cup Y) / (sup(X) \times sup(Y))$.
\end{itemize}



%
These metrics help identify meaningful rules; however, setting appropriate threshold values can be challenging.
ARM algorithms often generate a large number of rules~\cite{selvi2009automated}, which can overwhelm the analysis process.

Low thresholds may capture rare but important patterns, especially in critical datasets. 
For example, in a dataset of 1,000 crime incidents, a minimum support of 0.1 requires at least 100 matching incidents to retain a rule. 
Lowering the threshold increases coverage but also inflates the number of rules.
The best threshold values and choice of metrics vary by application. 
In this context, interactive visualization plays a key role. 
It allows analysts to explore different thresholds and evaluate rules dynamically.

Our approach builds on these foundations to support the analysis of spatiotemporal data with ARM. 
We utilize association rules to extract relevant patterns automatically. These patterns serve as guidelines that facilitate the visual exploration process. 
Unlike black-box models, association rules provide transparent and interpretable insights that are directly tied to the dataset's attributes.


\begin{figure*}[ht]
    \centering
    \includegraphics[width=1\linewidth]{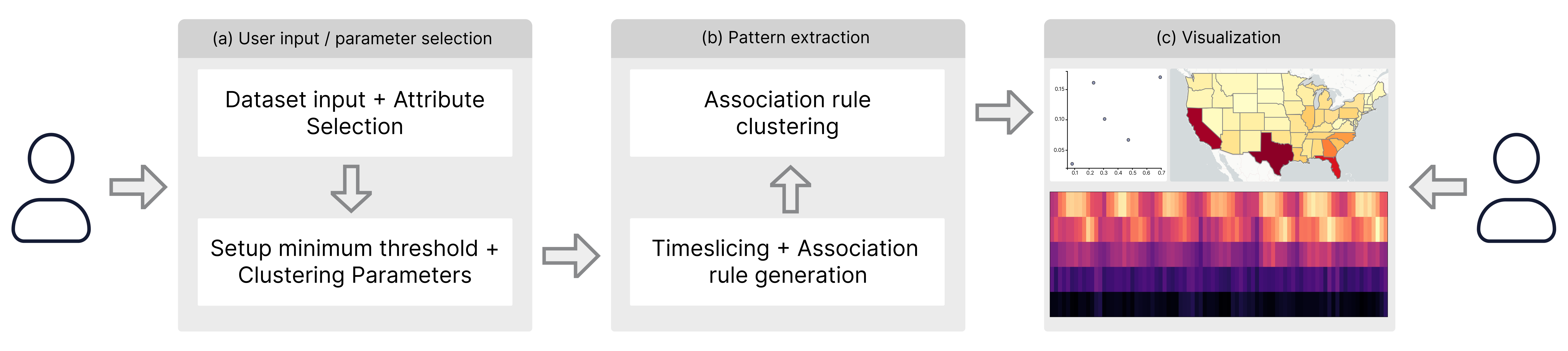}
    \caption{\systemname workflow. (a) The user provides a dataset of interest, selects relevant attributes, sets up minimum thresholds for rule generation, and defines clustering parameters. (b) The system generates rules and clusters based on the user's configuration and computes their metrics. (c) Finally, users explore the generated patterns through six visual components, organized into two views. }
    \label{fig:workflow}
\end{figure*}

\section{System Overview}
\label{sec:overview}
This section outlines the design methodology of our system. 
We describe the system requirements, analytical tasks, and overall workflow implemented in \systemname.

\subsection{System Requirements}
To guide users toward meaningful and explainable patterns, we define key design requirements for \systemname. 
These requirements are based on a review of interactive visual analytics for spatiotemporal pattern discovery, as well as our team's prior research experience.

\begin{itemize}[nosep,leftmargin=*]
    \item \textbf{Analyze relationships among attributes.} Understanding how attributes co-occur can uncover hidden patterns. Due to the numerous possible attribute combinations, users benefit from automated techniques that identify frequent associations. ARM supports this task by identifying recurring combinations in the dataset \cite{altaf2017applications, selvi2009automated}.
    \item \textbf{Identify temporal trends and spatial patterns.} Detecting how patterns evolve over time and space is essential for understanding trends and anomalies. Interactive visualizations help users observe when and where association rules appear \cite{fister2023comprehensive}.
    \item \textbf{Manage a large number of association rules.} ARM often produces many rules, making interpretation difficult. Clustering and summarization techniques simplify exploration and help users compare patterns more easily \cite{hahsler2017visualizing}.
\end{itemize}

The first two requirements align with long-established goals in spatiotemporal visualization research, whereas the third addresses specific limitations inherent to applying ARM in this context.

\subsection{Analytical Tasks}
\label{sec:tasks}

Based on the system requirements, we define the main analytical tasks supported by \systemname. 
These tasks help users discover patterns, observe temporal and spatial concentrations, detect shifts over time, and explore results interactively.

\begin{itemize}[nosep,leftmargin=*]
    \item \textbf{T1. Attribute, date, and algorithm configuration.} The system should offer controls to select attributes of interest, define a time range, set thresholds for rule measures (\eg, support, lift), and configure clustering granularity.
    \item \textbf{T2. Rule and cluster composition comparison.} The user should be able to compare common associations at the cluster level and explore attribute differences at the rule level in a unified view.
    \item \textbf{T3. Temporal behavior analysis.} The system should display the frequency of each rule and cluster over time slices. This helps detect recurring trends and unusual time-dependent behaviors.
    \item \textbf{T4. Spatial distribution analysis.} The user should be able to examine the spatial distribution of rule and cluster occurrences to understand where specific patterns emerge.
    \item \textbf{T5. Rule and cluster ordering.} The system should organize rules and clusters based on the similarity of their temporal patterns.
    \item \textbf{T6. Cluster metric display.} The system should present key cluster metrics\,---\,such as rule count, average support, and average confidence\,---\,in one layout to simplify performance comparison.
\end{itemize}

\subsection{Workflow}
\label{sec:workflow}

Figure~\ref{fig:workflow} illustrates the system workflow. Users begin by uploading a dataset in CSV format, which includes temporal and location fields, as described in Section~\ref{sec:background}. The tool accepts uploading datasets with multiple categorical attributes, provided they contain both a \texttt{DATE} column (formatted as \texttt{YYYY-mm-dd}) and a \texttt{PLACE} column. Users must also upload a GeoJSON file to render the locations specified in the \texttt{PLACE} field. The place names in the GeoJSON file must exactly match the values in the \texttt{PLACE} column so that aggregated data displays correctly. After both files are loaded, users select the attributes to be used for generating association rules.
%
%
Records with null or `Unknown' values in selected columns are discarded to prevent meaningless associations.
When attributes contain many distinct values, users should group similar values to reduce complexity. For instance, they can categorize vehicle models into broader types such as SUVs, sedans, or trucks. This improves rule generation by increasing support levels.
Next, users define the analysis time range and set thresholds for support and lift. 
They also configure a resolution parameter for clustering, explained in Section~\ref{sec:armcluster}.

The system then extracts and clusters association rules (Figure~\ref{fig:workflow}.b). 
It filters records by the selected time range and divides the data into time slices. 
For each slice, the system generates rules using the specified configuration. 
It removes duplicate rules that appear across slices and applies clustering to the combined rule set. 
Section~\ref{sec:armcluster} details the rule generation, deduplication, and clustering steps. 
Finally, users explore the patterns through \systemname's interface (Figure~\ref{fig:workflow}.c). 
The interface includes coordinated views for analyzing clusters and drilling down into individual rules. 
Section~\ref{sec:interface} describes the interface components and interactions.

\section{Association Rule Mining and Clustering}
\label{sec:armcluster}

A key feature of our system is the drill-down analysis capability. 
Users begin with an overview of rule clusters and then proceed to a detailed exploration of individual rules. 
This section describes the processes of generating and clustering association rules. 

Our system utilizes the FP-Growth algorithm for ARM due to its computational efficiency compared to alternatives, such as Apriori. 
Users typically set thresholds, such as minimum support and lift, to discard irrelevant rules.

Our primary goal is to identify spatiotemporal patterns using ARM. 
To incorporate temporal information, we divide the dataset into multiple time slices, as detailed in Section~\ref{sec:workflow}, and generate rules separately for each period. 
Generating rules for the entire dataset may highlight global trends but can miss important time-specific patterns. 
By examining rules across multiple slices, we track their evolution over time.

However, two key challenges can complicate the visualization and interpretation of these temporal rules. 
First, small temporal shifts in data can generate \textit{superfluous} rules\,---\,rules differing only by attribute placement between antecedent and consequent. For example, it is common to generate rules such as the following three:
\begin{align*}
    \{A, B, C\} \Rightarrow \{D\} \\
    \{A, B \} \Rightarrow \{C, D\}  \\
    \{A, B, D\} \Rightarrow \{C\} 
\end{align*}

\begin{figure*}[t!]
    \centering
    \includegraphics[width=1\linewidth]{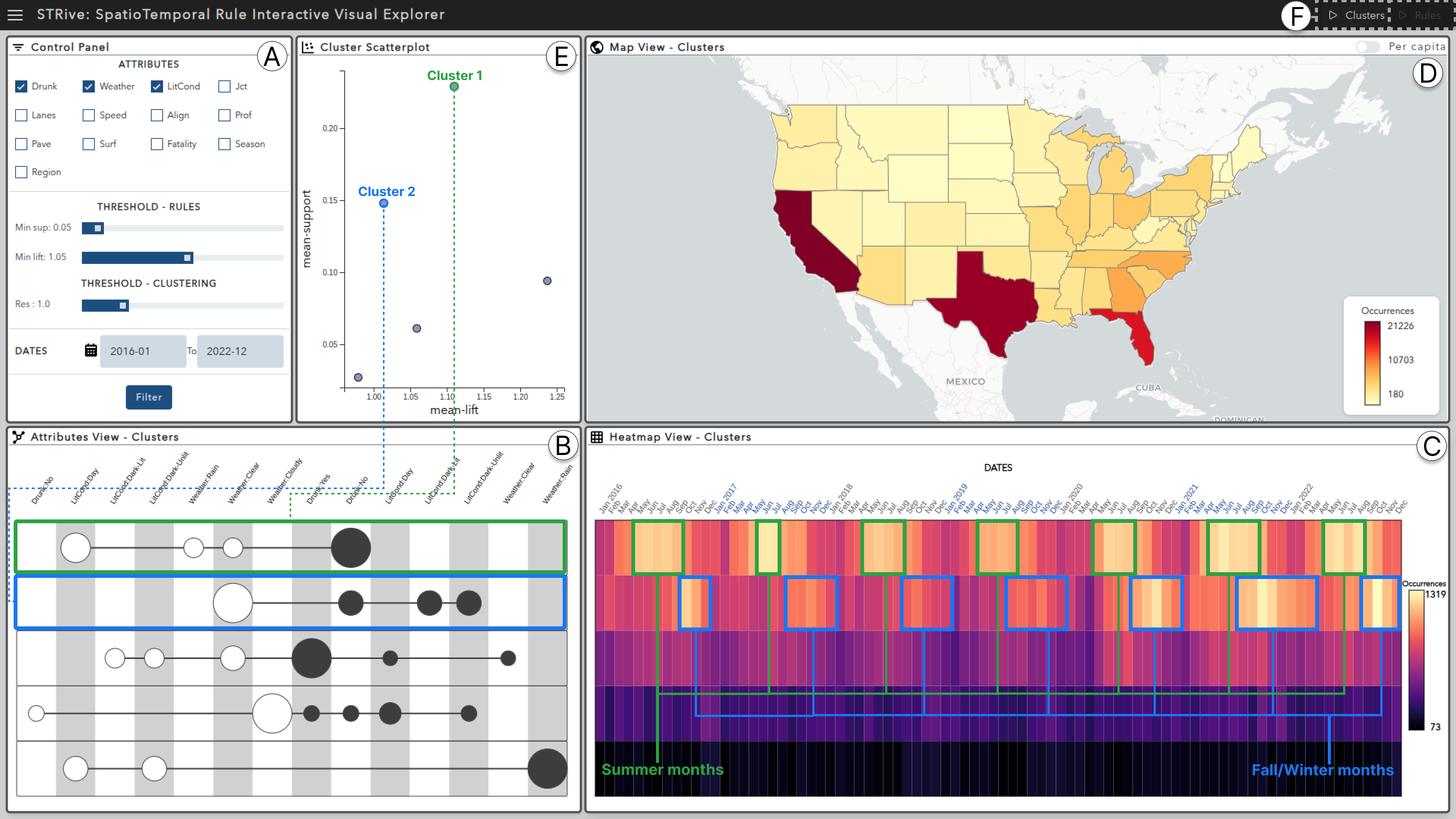}
    \caption{\systemname interface with annotated components from the first case study. (A) \emph{Control Panel} – allows users to select attributes and configure rule thresholds. (B) \emph{Attributes View} – displays the composition of each cluster. (C) \emph{Heatmap View} – reveals the temporal behavior of clusters across time slices. (D) \emph{Map View} – shows the spatial distribution of rule clusters. (E) \emph{Cluster Scatterplot} – provides a summary of cluster performance based on rule metrics. (F) \emph{Mode Selector} – allows to change between \emph{cluster-level} and \emph{rule-level} modes.} 
    \label{fig:interface}
\end{figure*}

\noindent Though semantically distinct, these rules share the same support. 
To reduce redundancy, we retain only the rule with the highest mean lift from each group of similar rules.
Second, even after this reduction step, the number of generated rules may remain large and difficult to visualize effectively. 
To handle this, we cluster rules into semantically coherent groups. 
We use the similarity metric from Fister~et~al.~\cite{fister2021association}, defined as:
\begin{equation*}
    \text{Sim}(r_{1}, r_{2}) = \frac{|\text{Ante}(r_{1})\cap\text{Ante}(r_{2})| + |\text{Cons}(r_{1})\cap\text{Cons}(r_{2})|}{|\text{Ante}(r_{1})\cup\text{Ante}(r_{2})| + |\text{Cons}(r_{1})\cup\text{Cons}(r_{2})|} \label{eq:sim}
\end{equation*}

This metric evaluates rule similarity based on common attributes in antecedents and consequents. 
We build a complete similarity graph, where nodes represent rules and edge weights reflect the degree of similarity between rules.
For clustering, we employ the Louvain algorithm~\cite{Blondel_2008}, a greedy, agglomerative method effective in detecting communities without predefined cluster counts. 
Preliminary tests confirmed its superior cluster coherence and computational efficiency performance. 
The Louvain algorithm includes a resolution parameter that controls cluster granularity: lower resolutions yield fewer, larger clusters, while higher resolutions create more numerous, smaller clusters. 
Users can adjust this parameter based on their analysis needs.
The resulting clusters serve as the basis for initial data exploration, providing meaningful groupings that simplify further analysis.

\section{The \systemname System}
\label{sec:interface}

\systemname is a novel visualization tool designed to support interactive visual analysis of spatiotemporal data via association rules. It comprises seven interactive components organized in two windows (the main interface is shown in Figure~\ref{fig:interface}) designed to support the design tasks described in Section~\ref{sec:tasks}. 
Some components support two modes, \emph{cluster-level} and \emph{rule-level}, which can be toggled using the buttons located in the top-right corner of the interface (Figure~\ref{fig:interface}.F). These modes adjust the views to accommodate different levels of detail. We describe each of these components in the rest of this section.

\myparagraph{Control Panel.}
The \emph{Control Panel} (Figure~\ref{fig:interface}.A) provides options for configuring the rule generation and clustering processes \textbf{(T1)}. Users can import a spatiotemporal dataset and select the relevant attributes to be considered for rule generation. Additionally, the panel allows the user to specify starting and ending dates for rule generation. To explore diverse clustering configurations, the panel includes an input field for adjusting the resolution parameter of the clustering algorithm, enabling users to control the size of the resulting rule clusters. 

\myparagraph{Attributes View.}
After users configure the rule-generation and clustering parameters in the \emph{Control Panel}, the resulting rule clusters appear in the \emph{Attributes View} (Figure \ref{fig:interface}.B). The design adapts the ARMatrix visualization by Varu et al.~\cite{varu2022armatrix} and extends the work of Valdivia et al.~\cite{valdivia2021analyzing} with additional information.
Each row represents one cluster \textbf{(T2, T5)}, and each column corresponds to an attribute–value pair found in the rules. Within a row, a circle marks the presence of an attribute; its radius encodes the frequency with which that attribute occurs across the cluster's rules, while its fill color distinguishes antecedent attributes (white) from consequent attributes (black).
In \emph{rule-level} mode, the view switches to display individual rules rather than clusters: each row now represents a single rule, the column layout is unchanged, and all circles have equal size because a given attribute can appear only once within a rule.
Users can interact with the \emph{Attributes View} by selecting a cluster, which updates the \emph{Map View} to display the occurrences satisfying the rules per location. Clicking on a specific cluster also makes the \emph{Attributes View} and \emph{Heatmap View} eligible for switching to \emph{rule-level} mode, allowing users to inspect the individual rules within that cluster if they choose to activate it.

\myparagraph{Heatmap View.}  
The \emph{Heatmap View} (Figure~\ref{fig:interface}.C) depicts the temporal evolution of the records that satisfy each rule cluster \textbf{(T3)}. Rows represent clusters, columns represent time intervals, and cell colour encodes the record count, making temporal trends easy to spot. Clusters with similar temporal signatures are placed adjacent to one another, and the row order is shared with the \emph{Attributes View}; scrolling in either view keeps the other in sync.
The view also supports a \emph{rule-level} mode. When a cluster is selected, each row switches to an individual rule while the columns continue to denote time intervals. The cell color now indicates how often that rule is satisfied in each period. This finer granularity complements the \emph{Attributes View} by directly linking rule structure to temporal behavior, while maintaining the same row order and linked scrolling for a cohesive exploration experience.

\myparagraph{Map View.}
The \emph{Map View} (Figure~\ref{fig:interface}.D) depicts the spatial distribution of records captured by the generated clusters \textbf{(T4)}. After rule generation and clustering, the map initializes at the spatial resolution defined by the dataset's geographic attribute. Each location is shaded with a sequential color scale that encodes the total number of occurrences observed at that location.
By default, the map aggregates occurrences from \emph{all} clusters, providing an overview of the entire dataset. Selecting a cluster in the \emph{Attributes View} filters the map to display only that cluster's occurrences, revealing its distinctive spatial footprint.
The view is tightly linked to the \emph{Heatmap View}: clicking a time label in the heatmap restricts the map to the chosen interval. If a cluster is selected, the filter becomes the intersection of the cluster and the period; otherwise, the map shows occurrences from every cluster within that period.
In \emph{rule‐level} mode, the design and interactions remain unchanged, but the focus shifts to individual rules. Selecting a rule in the \emph{Attributes View} updates the map to display its corresponding occurrences, and time filtering continues to operate via the heatmap, allowing users to track how each rule's spatial pattern evolves over time.

\begin{figure}[h]
    \centering
    \includegraphics[width=1\linewidth]{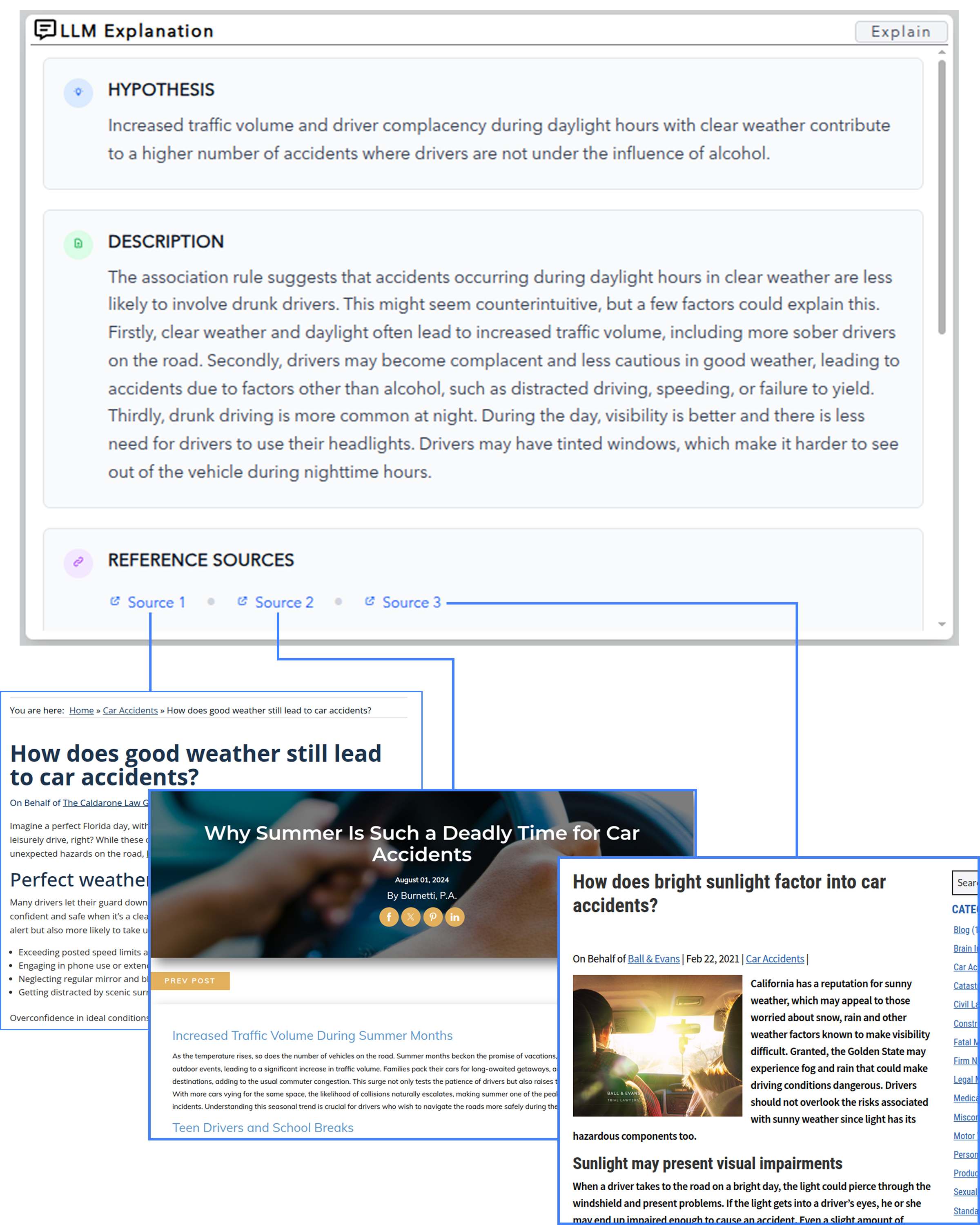}
    \caption{\systemname LLM explanation panel. The component uses the \texttt{gemini-2.0-flash} model to generate multiple hypotheses explaining the selected rule and searches the internet for supporting sources.}
    \label{fig:LLMexplain}
\end{figure}

\myparagraph{Cluster Scatterplot.}
To facilitate easy comparison of cluster performances, the \emph{Cluster Scatterplot} (Figure~\ref{fig:interface}.E) provides an overview of cluster-level metrics \textbf{(T6)}. Each dot in the scatterplot represents a cluster, and two metrics determine its position. By default, the X-axis shows the number of rules within a cluster, and the Y-axis displays the mean lift of the cluster. Customization of the scatterplot can be done by clicking on either axis label, which will display a dropdown menu listing additional metrics: mean number of occurrences, mean support, and mean confidence. Upon selecting a different metric, the scatterplot is redrawn to reflect the updated metrics. Additional interaction between views was implemented: selecting a cluster in the \emph{Attributes View} highlights its corresponding point in the scatterplot, while choosing a point in the scatterplot highlights the respective cluster in the \emph{Attributes View}. This enables the joint analysis of cluster composition and cluster metrics. This component is only available in cluster mode, as its purpose is to compare general properties across clusters.

\begin{figure*}[ht]
    \centering
    \includegraphics[width=1\linewidth]{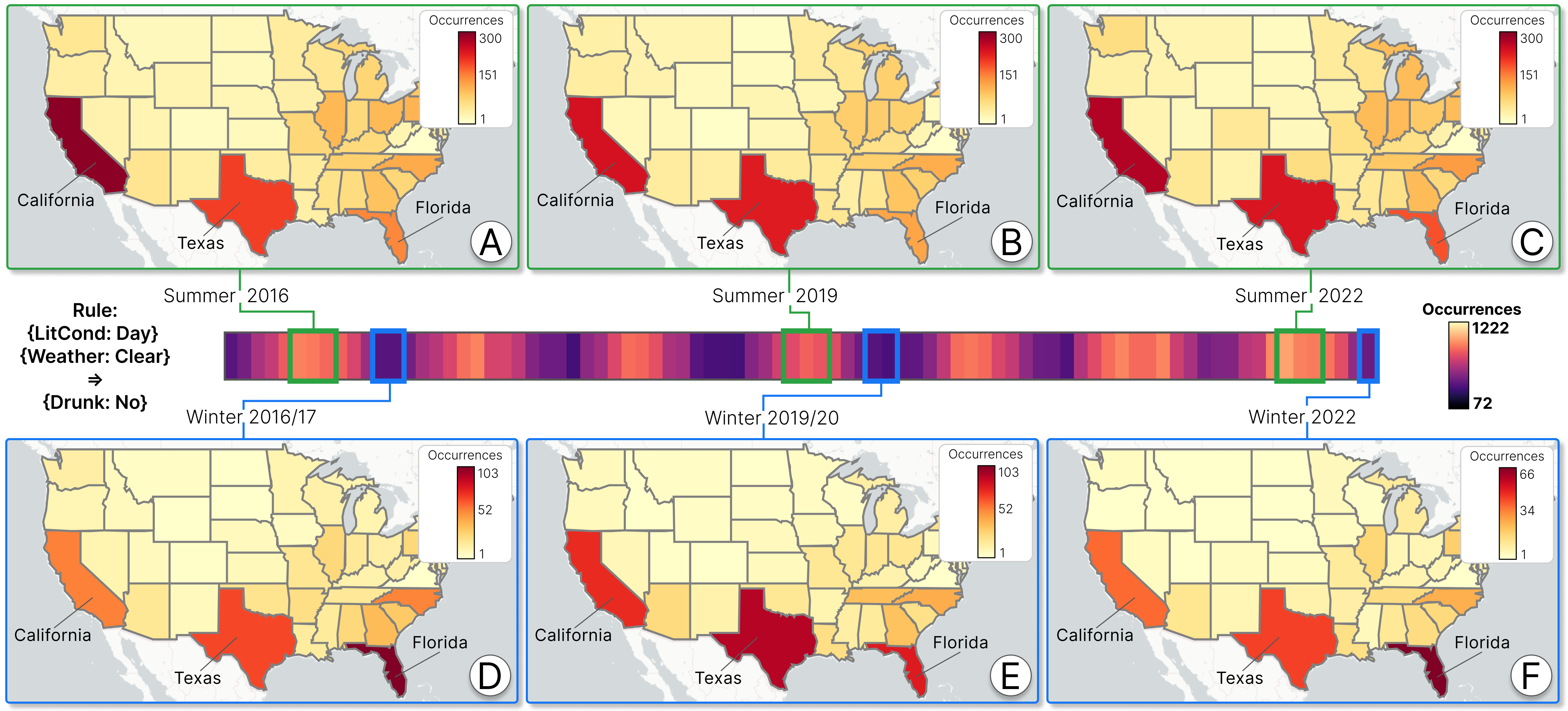}
    \caption{Example rule from Cluster 1 involving the attributes \texttt{Drunk}, \texttt{Lit Condition}, and \texttt{Weather}. During the summer months, accidents characterized by daylight, clear weather, and non-drunk drivers tend to concentrate in California and Texas (A, B, and C). In contrast, during the winter months, such accidents increase notably in Florida (D, E, and F).}
    \label{fig:fars1}
\end{figure*}

\myparagraph{LLM Explanation Panel.}
The \emph{LLM Explanation Panel} (Figure~\ref{fig:LLMexplain}) employs a large-language model (\texttt{gemini-2.0-flash}) to generate contextual explanations for a selected rule. After users select a rule in the \emph{Attributes View}, they can refine the scope by choosing specific regions on the map or particular time intervals. If no additional selection is made, the system defaults to the three locations with the highest rule frequency and the full date range. Pressing the \emph{Explain} button sends this context as a prompt (see \ref{prompt}.1) to the LLM, which returns hypotheses, pattern descriptions, and links to relevant external sources such as news articles. The resulting insights are displayed in the panel. This component is available only in \emph{rule-level} mode, where such fine-grained explanations are most useful.

\myparagraph{Implementation details}.
The system was implemented using Python, JavaScript, and HTML. The backend was built using Flask~\footnote{\url{https://flask.palletsprojects.com}}, the mlxtend~\footnote{\url{https://github.com/rasbt/mlxtend}} library provided the rule generation algorithms, and for clustering, the Louvain algorithm was implemented using the networkx library. On the client side, the interface was developed with HTML and JavaScript. The visualizations were implemented using D3.js~\footnote{\url{https://d3js.org}} and Leaflet~\footnote{\url{https://leafletjs.com}}.
The reorder.js library was used to reorder the rows in the \emph{Attributes View} and \emph{Heatmap View} components, based on the occurrences per time slice. 
In the present implementation, the algorithms for rule generation and clustering are fixed. Nevertheless, the preprocessing pipeline is modular, so replacing it is straightforward: users need only update the relevant module's function calls, provided the new algorithms emit output in the same format. All materials, including the source code, are available on our project website (\url{https://visualdslab.com/papers/STRive/}). This work has received the \textit{Replicability Stamp} (\url{https://www.replicabilitystamp.org/\#https-github-com-visual-ds-strive}), certifying that its results are reproducible.
\section{Case Studies}
\label{sec:cases}

We present two case studies to demonstrate \ systemname's capabilities for analyzing association rules.

\subsection{Analyzing Seasonal Patterns in Vehicular Accidents}
Our first case study explores factors related to fatal vehicular accidents. 
We use the Fatality Analysis Reporting System (FARS) dataset~\cite{FARS}, which contains detailed records of fatal accidents in the United States. 
Given the dataset's extensive nature, we select only relevant attributes and rename them for clarity. The list of attributes and their descriptions is provided in \ref{appendix:fars_data}.

We analyze data from 2016 to 2022. 
Our goal is to understand how weather, public illumination, and driver intoxication (attributes: \texttt{Weather}, \texttt{Lit Condition}, \texttt{Drunk}) relate to fatal accidents. 
After removing null and `Not Reported' values, we split the data into monthly segments, each with approximately 2,000 accidents.
We set the minimum support at 0.05, the minimum lift at 1.05, and the cluster resolution at ~1.
The resulting clusters contain 2, 4, and 6 rules. 
Figure~\ref{fig:interface}.B highlights the two clusters with more occurrences (Cluster 1 and Cluster 2). 
Cluster 1 involves clear weather (\texttt{Weather:Clear}), daylight (\texttt{LitCond:Day}), and non-drunk drivers (\texttt{Drunk:No}). 
Cluster 2 includes clear weather, dark conditions (\texttt{LitCond:Dark-Lit} and \texttt{LitCond:Dark-Unlit}), and non-drunk drivers (\texttt{Drunk:No}).

\begin{figure*}[ht]
    \centering
    \includegraphics[width=1\linewidth]{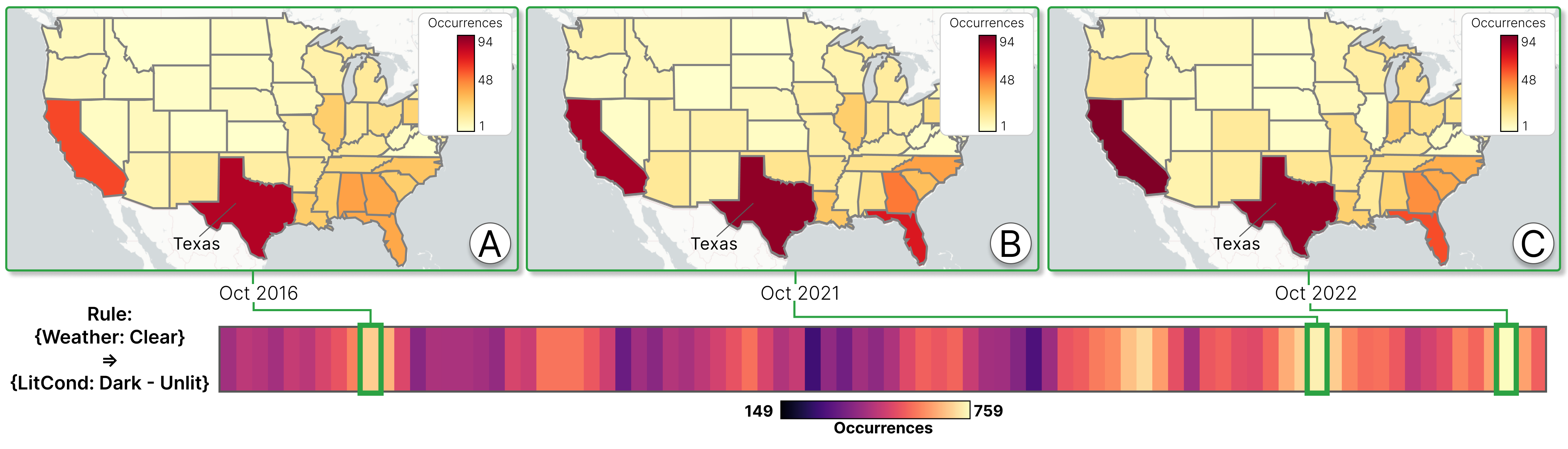}
    \caption{Rule from Cluster 2 describing nighttime accidents under unlit conditions. The state of Texas consistently appears as one of the most affected regions, with notable peaks in incident counts during October (A, B, and C).}
    \label{fig:fars2}
\end{figure*}

\begin{figure*}[t!]
    \centering
    \includegraphics[width=0.9\linewidth]{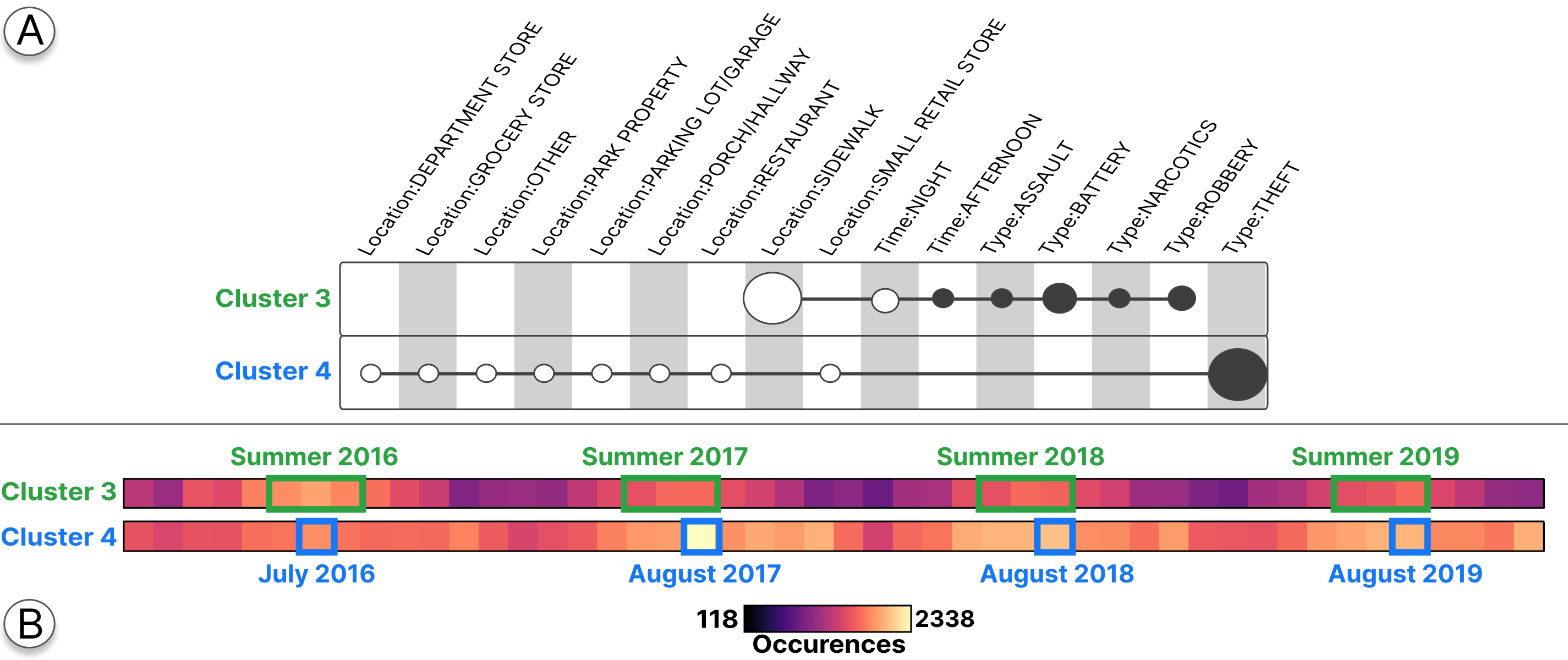}
    \caption{Two clusters selected for analysis. (A) Cluster composition. (B) Cluster temporal behavior.}
    \label{fig:chicago_general}
\end{figure*}

These associations help analyze the role of lighting conditions in fatal accidents. 
The \emph{Cluster Scatterplot} (Figure~\ref{fig:interface}.E) indicates both clusters have reliable associations (mean lift $>$ 1).
Cluster 1 has a higher average support than Cluster 2, meaning its described conditions occur more frequently.
The \emph{Heatmap View} (Figure~\ref{fig:interface}.C) shows monthly occurrences. 
Cluster 1 peaks from May to August, while Cluster 2 increases from October to January. 
Although Cluster 1 generally exhibits higher occurrence rates, both clusters exhibit clear seasonal peaks. 
The remaining rows show temporal behaviors involving drunk drivers (\texttt{Drunk:Yes}).
The significant difference between Clusters 1 and 2 is the lighting conditions. 
This suggests that lighting has a greater influence on fatal accidents than driver intoxication\,---\,the remaining clusters involving similar conditions and drunk drivers have fewer occurrences. 
We further analyze individual rules within these clusters.

This indicates a spatial shift in the distribution of rules.

Selecting Cluster 1, we examine daytime accidents (\texttt{Lit Condition:Daylight}). 
We set the tool to \emph{rule-level} mode by clicking on the \textit{Rules} button (Figure~\ref{fig:interface}, F). 
The components in the \textit{rule-level} mode mirror their counterparts in the \textit{cluster-level} mode, allowing us to visually identify the rule structure and inspect their temporal and spatial behavior.
Figure \ref{fig:fars1} illustrates the temporal pattern of the rule \{LitCond:Day, Weather:Clear\} $\Rightarrow$ \{Drunk:No\}, confirming a summer peak. 
Filtering data for June, July, and August, the \textit{Map View} reveals that most accidents occur in California and Texas (Figure~\ref{fig:fars1}.A–C). 
Conversely, selecting December and January shifts the highest incidence to Florida (Figure~\ref{fig:fars1}.D–F).
This shift aligns with reports indicating an increase in winter travel to Florida due to milder weather, which increases the flux of people traveling to the state, causing the number of vehicle accidents to rise~\cite{news1, news2, news3}.
Next, we explore nighttime accidents from Cluster 2.
Figure \ref{fig:fars2} shows the rule \{LitCond:Dark-Unlit, Weather:Clear\} $\Rightarrow$ \{Drunk:No\}. 
Accidents peak consistently in October. 
Spatial analysis reveals that Texas consistently has high accident rates this month (Figure~\ref{fig:fars2}.A–C). 
This observation aligns with official reports stating that October is particularly deadly for pedestrians in Texas due to reduced daylight hours~\cite {news4}.

\subsection{Analyzing Crimes in Chicago}
Our second case study explores factors related to crime incidents in Chicago. The data comes from the Chicago Police Department's CLEAR (Citizen Law Enforcement Analysis and Reporting) system~\cite{Chicago}. 
This dataset includes date, location, crime type, and arrest status. For our analysis, we select records from 2016 to 2019. 
We focus on examining the relationship between crime types and locations (attributes: \texttt{Type} and \texttt{Location}) and the timing of incidents (attribute: \texttt{Time}). 
After preprocessing to remove null and "Not Reported" entries, we segment data monthly, resulting in approximately 17,000 to 25,000 incidents per month.
Given the large dataset, we set minimum \textit{support} and \textit{lift} thresholds at 0.01 and 1.5, respectively, to capture meaningful associations. 
We also set the clustering resolution parameter at 2.5 to achieve more refined clusters.

\begin{figure}[t!]
    \centering
    \includegraphics[width=1\linewidth]{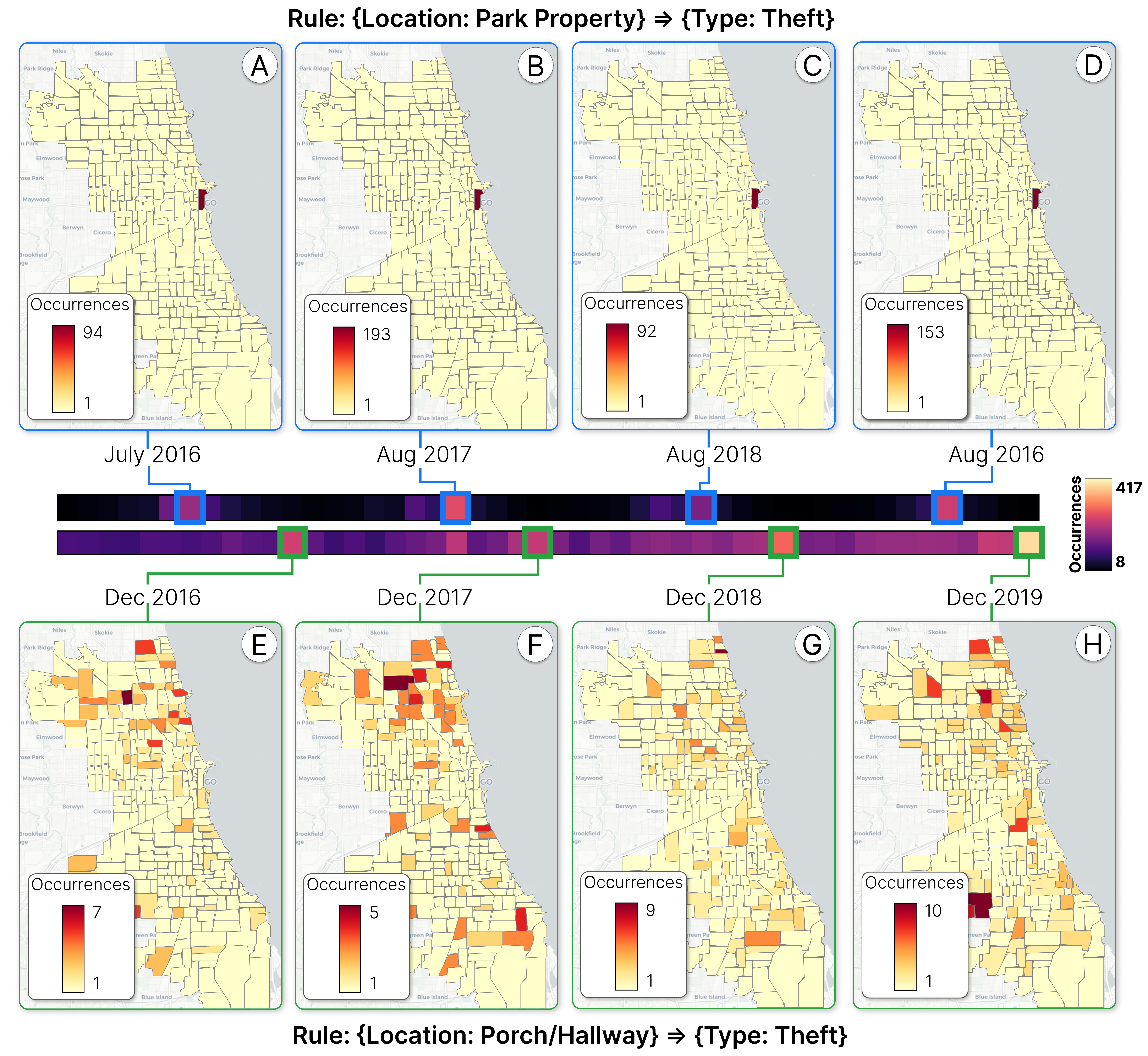}
    \caption{Two example rules from Cluster 4. The top row (A, B, C, and D) illustrates the spatial distribution of thefts that occur in parks. The bottom row (E, F, G, and H) shows the spatial distribution of thefts on residential porches and in hallways.}
    \label{fig:chicago1}
\end{figure}

Figure~\ref{fig:chicago_general}.A displays the resulting clusters. 
Our analysis focuses on two clusters, specifically Cluster 3 and Cluster 4. 
Cluster 3 includes rules involving multiple crime types that occur primarily on sidewalks (\texttt{Location:SIDEWALK}), mostly during the afternoon (\texttt{Time:afternoon}) and nighttime hours (\texttt{Time:night}). 
Cluster 4 consists exclusively of theft incidents (\texttt{Type:THEFT}) across various locations (see attribute \texttt{Locations}. 
These clusters stand out due to their distinct seasonal trends (see Figure \ref{fig:chicago_general}.B). 
Cluster 3 has the highest number of incidents and exhibits clear seasonality, with peaks in the summer. 
Cluster 4 also demonstrates seasonality, though weaker, with notable spikes.

\begin{figure}[t!]
    \centering
    \includegraphics[width=1\linewidth]{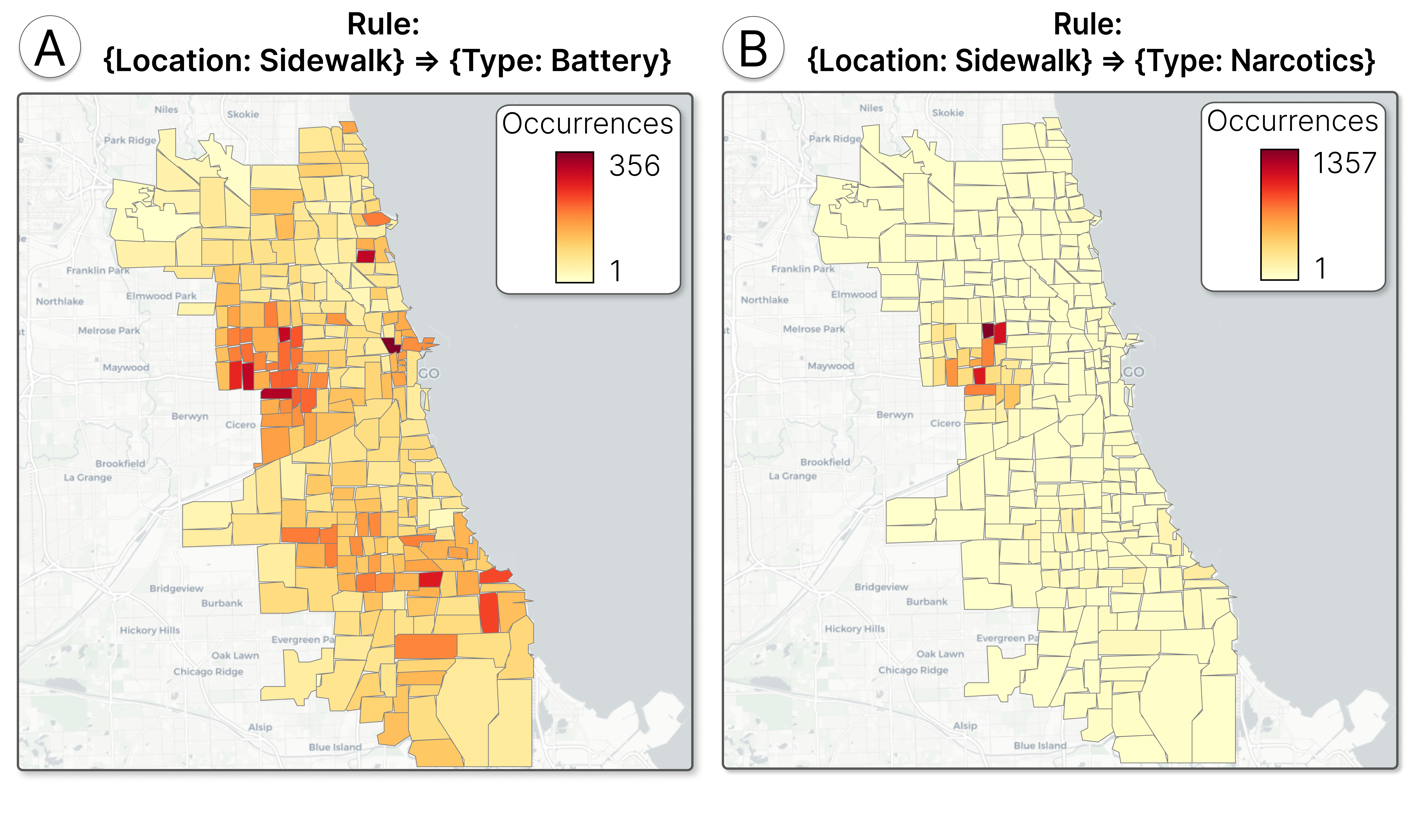}
    \caption{Spatial distribution of different crime types. (A) Map showing the distribution of battery incidents. (B) Map showing the distribution of narcotics-related offenses.}
    \label{fig:chicago2}
\end{figure}

We first examine Cluster 4 to understand the seasonal spikes. 
The cluster contains rules describing thefts at locations such as parks, residential porches and hallways, and grocery stores (Figure~\ref{fig:chicago_general}.A).
A detailed temporal analysis (Figure~\ref{fig:chicago1}) reveals that thefts on porches and hallways peak during December and January, likely due to increased online shopping and unattended deliveries during the holidays.
Conversely, thefts in parks spike notably in July and August, despite being relatively rare. 
Spatial analysis identifies the Loop district, specifically Beat 0114, as the primary location for these park thefts (Figure~\ref{fig:chicago1}.A–D). 
Grant Park, within Beat 0114, hosts the Lollapalooza festival during peak theft periods, explaining the spikes in theft incidents during these months (July 28–31, 2016; August 3–6, 2017; August 2–5, 2018; August 1–4, 2019).

Next, we examine the rules in Cluster 3, with a focus on sidewalk-related incidents. 
The cluster prominently features battery incidents on sidewalks (see rule on Figure~\ref{fig:chicago2}.A). 
Examining narcotics offenses on sidewalks reveals high incident concentrations in West Garfield Park, Humboldt Park, and Austin (see rule on Figure~\ref{fig:chicago2}.B). 
These areas are known for gang activity and drug-related issues, aligning with reports indicating they account for approximately 30\% of Chicago's emergency drug-related calls~\cite{news5}.
The different treatments of time and location in the two case studies stem from differences in the underlying datasets. In the crime dataset, location is already provided as a categorical attribute (e.g., "sidewalk"), and time can be discretized into broad periods such as morning, afternoon, and evening. Both attributes can therefore be included in rule generation. By contrast, the vehicular-accident dataset offers richer contextual variables\,—\,such as speed limit, road type, and light condition\,—\,while its geographic information is limited to continuous coordinates, which are not suitable for direct ARM. Consequently, the vehicular accident case study focuses on those categorical road context attributes instead. In both studies, spatial and temporal information is employed mainly in the analysis phase to contextualize the resulting rules.
\section{Expert Feedback}

To further evaluate our system, we conducted semi-structured interviews with domain experts, focusing on its usability and usefulness. This section describes the participants' backgrounds, the interview methodology, and the main findings.

\subsection{Participants’ Backgrounds}

We invited two computer science experts, each with over 20 years of experience in data analysis. One specializes in data visualization, and the other in data mining; both have extensive hands-on experience with ARM in their research. Their combined expertise, therefore, spans the principal areas on which our system is built.


\subsection{Interview Process}

We conducted separate one-hour semi-structured interviews with each expert.  
We began by introducing the class of datasets targeted by STRive and outlining the main steps of the methodology—rule generation, rule clustering, and metric computation.  
Next, we demonstrated the interface in detail, explaining how each component supports insight discovery.  
We then presented our case studies and provided a link to a live demo populated with the same data, allowing the experts to explore the tool hands-on.  
Finally, each expert completed a questionnaire that captured their views on the methodology, the system's usefulness and usability, and potential areas for improvement.

\subsection{Results}
Below, we summarize the main feedback obtained during the interviews.

\myparagraph{Methodology.}  
When asked, \emph{``What is your opinion of the proposed methodology for mining spatiotemporal patterns?"} Both experts described the approach as compelling and pointed to the accompanying use cases as persuasive examples. One expert observed that the current framework could not capture more complex temporal structures and recommended extending it to support sequential rule mining~\cite{huang2008framework}. The other expert noted that analysts rarely begin exploration with the entire rule set produced by an ARM algorithm; instead, they typically start with a specific hypothesis\,---\,for example, requiring a given attribute to appear in the consequent\,---\,and then investigate the factors that lead to that outcome. Although STRive does not yet support this workflow, it could be implemented by filtering the generated rules for user-specified attribute values. We plan to incorporate this capability in the next version of the system.

\myparagraph{Usability.}  
In response to the questions \emph{``Was the system easy to use?"} and \emph{``Do you believe people will learn how to use this system quickly?"} Both experts reported that the interface is straightforward and the visualizations are intuitive. They cautioned, however, that their considerable experience with association-rule mining may influence this impression\,---\,and, in one case, specifically with ARM visualization.
When asked whether setting the initial parameters or selecting attributes for rule generation was difficult, both replied that the process was simple. One expert noted that the evaluation datasets contained relatively few attributes, making the checkbox selection effective; with larger datasets, the same approach might become a usability bottleneck.

\myparagraph{Usefulness.}  
In response to \emph{``Do you consider STRive a useful tool, and why?"} Both experts affirmed its utility. One highlighted that STRive is especially compelling when patterns emerge in specific regions during specific time frames, where it could support decision-making\,---\,for instance, helping policymakers address pedestrian incidents in October in Texas. He remarked: \emph{``The main benefit I see is STRive's capacity to abstract data into simple, interpretable patterns that are otherwise very difficult to extract. In scenarios where such abstraction is acceptable, the tool can indeed be very useful.''}

The second expert echoed this view and emphasized the value of visualizing incidents that satisfy the rules across regions and over time. He expressed interest in being able to switch the displayed metrics (e.g., showing \emph{lift} instead of the number of occurrences) in both the \textit{Heatmap} and \textit{Map views}, as well as in supporting multiple temporal resolutions (e.g., weekly) in the \textit{Heatmap}.

Both experts felt that all components are important for the analyses shown in the case studies; none was judged more critical than the others. They specifically praised the LLM explanation panel, noting that it substantially enhances the interpretability of the discovered patterns.

\myparagraph{Limitations.}  
Asked about potential shortcomings, the first expert observed that datasets with hundreds of attributes—common in spatiotemporal analysis—could overwhelm the interface and that rules containing many attributes may impose a high cognitive load. He stressed, however, that these challenges are intrinsic to association-rule mining rather than unique to STRive. The second expert reiterated scalability concerns and pointed out the potential loss of interpretability when clustering rules. He acknowledged, though, that clustering is necessary to manage the large rule sets and to enable focused, drill-down exploration.

Overall, the expert evaluation confirmed STRive's ease of use and practical value while highlighting its limitations and suggesting directions for future improvement.

\section{Discussion and Future Work}
\label{sec:discussion}

As shown in Section~\ref{sec:cases}, \systemname enables users to identify and analyze spatiotemporal data patterns effectively. 
This section discusses the current workflow, limitations, and future directions for improving \systemname.

\myparagraph{Use of clustering algorithms.}
We currently use the Louvain algorithm to create coherent rule clusters. 
After evaluating multiple methods, Louvain demonstrated superior results in terms of cluster quality, parameter flexibility, and computational efficiency. 
However, Louvain is non-deterministic, meaning different runs may produce varying clusters. 
As the clustering step plays a major role in our workflow, this variability can hinder reproducibility and complicate analyses. 
Future implementations could include alternative clustering methods with user-defined parameters. 
Additionally, extending clustering to incorporate spatial and temporal rule distributions could provide more comprehensive insights.

\myparagraph{Interactive rule modification.} 
Currently, \systemname supports rule comparison based on spatial and temporal attributes. 
Allowing users to add or remove attributes within rules dynamically would provide more direct rule manipulation. 
This feature could guide initial exploration and reveal additional patterns derived from user-modified rules.

\myparagraph{Computational Scalability.} 
Our current implementation can handle moderately sized datasets with diverse spatiotemporal attributes; however, it is not specifically optimized for ``big-data'' workloads. Computational bottlenecks primarily arise in the rule-extraction and clustering phases. These bottlenecks can be mitigated in several ways, as both tasks are well-studied. For instance, more sophisticated algorithms for rule mining and clustering could significantly improve STRive's scalability~\cite{zhao2021improved,wang2021research,do2024improvement}.

\myparagraph{Visual Scalability.} The present implementation of the \textit{Heatmap} and \textit{Attributes views} does not scale well as the number of time slices or data attributes grows. This limitation can be alleviated by incorporating filtering\,---\,for example, focusing on specific periods or attribute subsets\,---\,and by adopting focus–plus–context techniques~\cite{cuenca2018multistream}. We plan to integrate these strategies in the next version of the system.

\begin{chatbox}
\label{promptdiscussion}
\textbf{User:} There was an increase in thefts in parks on the following dates: July 2016, August 2017, August 2018, and August 2019 in the Loop district in Chicago. Do you know any possible causes?
\newline
\newline
\textbf{ChatGPT:} The increases in thefts in Chicago's Loop district parks during July 2016, August 2017, August 2018, and August 2019 can be attributed to several factors:

1. Large Summer Events: Major festivals, such as Lollapalooza, held annually in late July or early August in Grant Park, draw substantial crowds...

2. Seasonal Crime Patterns: Warmer summer months often correlate with increased crime rates...

3. Resource Allocation Challenges: Large events require significant police presence, which can strain resources and potentially leave other areas more vulnerable...

4. Public Transit Crime: There have been instances of increased crime on Chicago's public transit during summer months...

These factors combined likely contributed to the observed increases in park thefts during the specified periods.
\newline
\newline
\textbf{User:} Can you provide the exact dates and locations of the Lollapalooza festival?
\newline
\newline
\textbf{ChatGPT:} Lollapalooza is an annual music festival held in Chicago's Grant Park. Here are the exact dates for the years you inquired about:

\begin{itemize}
    \item 2016: July 28–31.
    \item 2017: August 3–6.
    \item 2018: August 2–5.
    \item 2019: August 1–4.
\end{itemize}

Each year, the festival spans four days, attracting large crowds to Grant Park in Chicago, Illinois.
\end{chatbox}

\myparagraph{Visual Encoding Enhancements.} Further work on visual encodings may help improve spatial interpretation by using glyphs to show the distribution of attributes within each region on the map. This could make it easier to spot local variations that are not visible through aggregate values alone.

\myparagraph{Pattern explanation support.} 
While our system identifies significant patterns, fully understanding their underlying causes often requires external context. 
Leveraging large language models (LLMs), such as ChatGPT-4o, proved beneficial for obtaining explanations in our first case study. 
For example, we asked about theft increases in Chicago parks during specific dates. 
ChatGPT-4o effectively identified factors such as major events and seasonal crime patterns contributing to this phenomenon. 
A partial interaction transcript appears in Conversation \ref{promptdiscussion}.1.
While \systemname already utilizes LLMs to generate contextual explanations and surface relevant external information, future enhancements could offer even more sophisticated guidance\,---\,for instance, helping analysts interpret visualizations and automatically highlighting potential patterns revealed by the system.

\myparagraph{Rule causality.} Although association rules can appear to express causal relationships, they in fact capture only frequent co-occurrences in the data. This limitation is inherent to ARM, so additional analysis—such as the interactive exploration offered by our system—is essential for identifying the rules that matter most. Users can refine the search by specifying attributes that must appear in the antecedent, the consequent, or both, thereby steering the mining process toward patterns that align with domain knowledge. Moreover, rule directionality plays a minor role in our workflow because we use \emph{lift}\,—\,a symmetric metric independent of rule direction\,—\,as the principal filtering criterion. 
%

\section{Conclusion}
\label{sec:conclusion}

This paper introduces \systemname, a visual analytics system that utilizes Association Rule Mining (ARM) as a guidance mechanism for exploring spatiotemporal data. \systemname leverages the FP-Growth algorithm to extract association rules and groups these rules into coherent clusters using the Louvain algorithm. The system provides users with interactive linked visualizations by computing rule metrics over time and spatial distributions.

Our work demonstrates that ARM effectively reveals meaningful patterns, significantly aids users in navigating complex spatiotemporal datasets, and inherits the ease of interpretation of these patterns. This feature is very much required in modern systems. The case studies validate the practical usefulness of ARM-based guidance, illustrating its ability to uncover insightful and interpretable relationships in real-world scenarios.

In future work, we plan to conduct a user study to evaluate the usability and effectiveness of \systemname formally. Additionally, we aim to integrate large language models (LLMs) to automatically generate explanations for the discovered patterns, further enhancing the system's interpretability and utility.
Finally, our system focuses on spatiotemporal \emph{event} data~\cite{kisilevich2010spatio}. Other forms (such as trajectories) pose different analytical tasks and pattern types, which may call for sequential or trajectory-aware rule-mining techniques~\cite{huang2008framework}. Extending \systemname\ to support these data, therefore, represents a promising avenue for future research.

\section*{Acknowledgments}
This work was supported by FAPESP (\#2021/07012-0), FAPERJ (\#E-26/204.593/2024, \#E-26/210.585/2025), CNPq (\#311144/2022-5, \#132348/2025-0, \#132349/2025-6, \#311425/2023-2), PROCIENCIA (\#PE501087483-2024), and Fundação Getulio Vargas. The opinions, hypotheses, conclusions, and recommendations expressed in this material are the responsibility of the authors and do not necessarily reflect the views of FAPESP,  FAPERJ, CNPq, PROCIENCIA, and Fundação Getulio Vargas.
\bibliographystyle{cag-num-names}
\bibliography{refs}

\newpage
\newpage
\appendix

\section{Fatality Analysis Reporting System (FARS) dataset}
\label{appendix:fars_data}
The FARS dataset includes a wide range of attributes related to different aspects of fatal vehicle accidents. In our analysis, we retain only a small subset of these attributes. Table \ref{table:fars-attr} presents the selected attributes and their descriptions. 

\begin{table}[h]
\centering
\resizebox{0.50\textwidth}{!}{
\begin{tabular}{|c|c|}
\hline
Attribute & Description \\ \hline
Id & Unique identifier for the accident. \\ \hline
Date & Date of the accident. \\ \hline
State & State where the accident happened. \\ \hline
Drunk & \begin{tabular}[c]{@{}c@{}}Indicates whether one of the involved\\ drivers was drunk.\end{tabular} \\ \hline
Lanes & \begin{tabular}[c]{@{}c@{}}Number of lanes of the road where the\\ accident occurred.\end{tabular} \\ \hline
Speed & \begin{tabular}[c]{@{}c@{}}Maximum speed limit of the road where\\ the accident occurred.\end{tabular} \\ \hline
Surface & \begin{tabular}[c]{@{}c@{}}Status of the road surface at the time\\ of the accident (e.g. Dry or Wet).\end{tabular} \\ \hline
Lit Condition & \begin{tabular}[c]{@{}c@{}}Status of the public illumination at the\\ time of the accident (e.g. Dark - Lighted,\\ Dark - Not Lighted or Daylight).\end{tabular} \\ \hline
Pavement & \begin{tabular}[c]{@{}c@{}}Type of pavement of the road where the\\ accident occurred (e.g. Asphalt).\end{tabular} \\ \hline
Level & \begin{tabular}[c]{@{}c@{}}Inclination of the road where the\\ accident occurred (e.g. Level or Uphill).\end{tabular} \\ \hline
Align & \begin{tabular}[c]{@{}c@{}}Alignment of the road where the\\ accident occurred (e.g. Straight or Curve).\end{tabular} \\ \hline
Junction & \begin{tabular}[c]{@{}c@{}}Accident location in relation to \\  its proximity to junction or interchange \\areas (e.g. Intersection or Driveway Access).\end{tabular} \\ \hline
Weather & \begin{tabular}[c]{@{}c@{}}Atmospheric conditions at the \\ time of the accident (e.g. Clear or Rain).\end{tabular} \\ \hline
Season & \begin{tabular}[c]{@{}c@{}}Season of the year when the accident \\ occurred (e.g. Summer or Winter).\end{tabular} \\ \hline
Region & \begin{tabular}[c]{@{}c@{}}Region where the accident  \\ occurred (e.g. Midwest or South).\end{tabular} \\ \hline
\end{tabular}
}
\caption{Attributes found in the FARS dataset.}
\label{table:fars-attr}
\end{table}

It is important to note that attributes such as Season and Region are not originally present in the data; they were derived from the Date and State attributes and added by us.
\section{Prompt Used for LLM Explanations}
\label{appendix:prompt}

\ref{prompt}.1 outlines the prompt used to generate explanations. Note that the text in \textbf{bold} highlights elements specific to the dataset in use. This prompt differs from the one in the Discussion section, as it is automatically generated from raw data, whereas the other was manually written after identifying the main pattern.

\begin{chatboxx}
\label{prompt}
Here we have an association rule, describing a pattern found in a \textbf{vehicular accidents dataset}. Each element in the antecedent or consequent is an attribute–value pair.
\\
\\
Antecedent: \kern-0.01em\{\textbf{LitCond:Daylight}, \textbf{Weather:Clear}\}

Consequent: \{\textbf{Drunk:No}\}
\\
\\
\textbf{Location:Florida}

\textbf{2016-01: 27}

\textbf{2016-02: 44}

\[ \vdots \] 
\textbf{Location:Texas}

\textbf{2016-01: 61}

\textbf{2016-02: 52}

\[ \vdots \]

Tasks:

1. Identify trends both in time and space.

2. Formulate a couple specific hypotheses explaining the identified behavior.

3. Search the internet for information sources to validate your hypothesis. 

3. Use the Google Search tool to find specific news articles, reports, and studies

4. Provide actual working URLs, not placeholder URLs.
            
If no information was found, just return the hypothesis and description.

Output the findings as a JSON list of dictionaries with the following format (strictly valid JSON only):
\begin{verbatim}
{
    "hypothesis": "",
    "description": "",
    "sources": []s
}
\end{verbatim}
Output each source as a JSON dictionary with title and URL:
\begin{verbatim}
{
    "title": "",   
    "url": ""
}
\end{verbatim}
\end{chatboxx}



\end{document}